\documentclass[aps,pra,superscriptaddress,floatfix,twocolumn,showpacs]{revtex4-1}
\usepackage{verbatim}
\usepackage{amsmath}
\usepackage{amsfonts}
\usepackage{amssymb}
\usepackage{graphicx}
\usepackage{color}
\usepackage{bm}
\usepackage{hyperref}
\usepackage{cases}
\usepackage{ulem}
\usepackage{multirow}
\usepackage{braket}

\newcommand{\naiseki}[2]{\left\langle #1 | #2 \right\rangle}

\begin{document}
\title{Bulk-edge correspondence in nonunitary Floquet systems with chiral symmetry}

\author{Ken Mochizuki}
\affiliation{Department of Applied Physics, Hokkaido University, Sapporo 060-8628, Japan}

\author{Dakyeong Kim}
\affiliation{Department of Applied Physics, Hokkaido University, Sapporo 060-8628, Japan}

\author{Norio Kawakami}
\affiliation{Department of Physics, Kyoto University, Kyoto 606-8502, Japan}

\author{Hideaki Obuse}
\affiliation{Department of Applied Physics, Hokkaido University, Sapporo 060-8628, Japan}

\begin{abstract}
We study topological phases in one-dimensional open Floquet systems driven by chiral symmetric nonunitary time evolution. We derive a procedure to calculate topological numbers from nonunitary time-evolution operators with chiral symmetry. While the procedure has been applied to open Floquet systems described by nonunitary time-evolution operators, we give the microscopic foundation and clarify its validity for the first time. We construct a model of chiral symmetric nonunitary quantum walks classified into class BDI$^\dagger$ or AIII, which is one of enlarged symmetry classes for topological phases in open systems, based on experiments of discrete-time quantum walks. Then, we confirm that the topological numbers obtained from the derived procedure give correct predictions of the emergent edge states. We also show that the model retains $\mathcal{PT}$ symmetry in certain cases and its dynamics is crucially affected by the presence or absence of $\mathcal{PT}$ symmetry.
\end{abstract}


\maketitle

\section{introduction}
\label{sec:introduction}
Understanding and controlling open systems are fundamentally important problems to be solved. Recently, the study on non-Hermitian Hamiltonians attracts great attention from various fields of physics, since non-Hermitian Hamiltonians can effectively describe open systems \cite{hatano1996localization,
bender1998real,
bender2002complex,
mostafazadeh2002pseudoI,
mostafazadeh2002pseudoII,
mostafazadeh2002pseudoIII,
mostafazadeh2003exact,
brody2013biorthogonal,
lee2014local,
tang2016experimental,
ashida2017parity,
ruter2010observation,
lin2011unidirectional,
miri2012large,
mostafazadeh2013invisibility,
guo2009observation,
peng2014loss,
peng2014parity,
feng2014single,
hodaei2014parity,
garmon2015bound,
longhi2018parity,
el2018non}. Especially, topological phases of such open systems have been intensively studied and unique features have been revealed \cite{esaki2011edge,
hu2011absence,
malzard2015topologically,
lee2016anomalous,
yuce2016majorana,
ling2016anomalous,
klett2017relation,
schomerus2013topologically,
poli2015selective,
weimann2017topologically,
weimann2017topologically,
bandres2018topological,
parto2018edge,
xiao2018higher,
gong2018topological,
kawabata2018parity,
kawabata2019symmetry,
ghatak2019new,
xiao2020non,
weidemann2020topological}. Non-Hermitian open systems have richer topological structures in comparison to Hermitian closed systems because the number of symmetries increases due to the absence of Hermiticity \cite{kawabata2019symmetry} and it depends on symmetries of systems whether non-trivial topological phases can exist or not \cite{schnyder2008classification,
kitaev2009periodic,
ryu2010topological}. There are various optical experiments in which non-Hermitian topological phases are explored and the associated edge states are observed \cite{poli2015selective,
zeuner2015observation,
xiao2017observation,
weimann2017topologically,
bandres2018topological,
parto2018edge,
xiao2018higher,
xiao2020non,
weidemann2020topological}. 

Among them, photonic quantum walks with effects of gain and/or loss provide ideal experimental platforms which are described by nonunitary time-evolution operators. This is because it is possible to tune various parameters of systems and experiments can be carried out in both classical \cite{regensburger2012parity,
regensburger2013observation,
wimmer2015observation} and quantum \cite{xiao2017observation,xiao2018higher,xiao2020non} regimes. Since quantum walks are Floquet systems in which time evolves in a discrete manner, topological phases can be different from those which are described by time-independent Hamiltonians. Floquet topological phases of quantum walks have been intensively studied for the last decade \cite{kitagawa2010exploring,
kitagawa2012topological,
obuse2011topological,
asboth2012symmetries,
asboth2013bulk,
asboth2014chiral,
edge2015lozalization,
asboth2015edge,
obuse2015unveiling,
cedzich2016bulk,
cardano2017detection,
endo2017sensitivity,
ramasesh2017direct,
verga2018entanglement,
barkhofen2018supersymmetric,
mochizuki2020stability,
mochizuki2020topological,
wang2020robustness,
xie2020topological}, and topological edge states have been observed in experiments of both closed \cite{kitagawa2012observation,
barkhofen2017measuring,
mukherjee2017experimental,
chen2018observation} and open \cite{xiao2017observation,
xiao2018higher,
chen2018characterization,
kawasaki2020bulk} systems. Especially, much attention has been paid to Floquet systems with chiral symmetry. This is because a procedure to calculate topological numbers has been established in the case of chiral symmetric unitary time-evolution operators \cite{asboth2013bulk}. On the other hand, regarding open Floquet systems described by nonunitary time-evolution operators, the microscopic foundation for the procedure has not yet been clarified although it has already been utilized in previous experimental and numerical studies \cite{xiao2017observation,
xiao2018higher,
chen2018characterization,
kawasaki2020bulk}. Also, the enriched symmetries of non-Hermitian open systems \cite{kawabata2019symmetry} have not been discussed so much in nonunitary open Floquet systems. \\\indent
In this work, we study topological phases and the corresponding edge states of chiral symmetric open Floquet systems with gain and loss in one dimension. We derive a procedure to calculate topological numbers for Floquet topological phases driven by chiral symmetric nonunitary time evolution based on discussions about the bulk-edge correspondence. While a method to calculate topological numbers in chiral symmetric unitary Floquet systems was obtained in Ref. \cite{asboth2013bulk}, we generalize the procedure to nonunitary chiral Floquet systems. We confirm the validity of the bulk-edge correspondence by using the derived topological numbers, for two different symmetry classes BDI$^\dagger$ and AIII in the classification of non-Hermitian topological phases proposed in Ref. \cite{kawabata2019symmetry}. To this end, we construct a chiral symmetric model of nonunitary quantum walks, whose setting is based on the experiments in Refs.\ \cite{regensburger2012parity,xiao2017observation}.  We find that the model also has $\mathcal{PT}$ symmetry in a specific case. While it is not related to $\mathcal{PT}$ symmetry whether topological edge states exist or not, we show that $\mathcal{PT}$ symmetry plays an important role in the dynamics of the nonunitary quantum walk with edge states. \\\indent
The rest of this paper is organized as follows. In Sec. \ref{sec:topological-numbers}, we derive topological numbers for chiral symmetric open Floquet systems. In Sec. \ref{sec:model}, we introduce a model of nonunitary quantum walks. We show symmetries of the model, and give a condition for the existence of quasienergy gaps. In Sec. \ref{sec:bulk-edge_correspondence}, we calculate topological numbers of the model by using the method derived in Sec. \ref{sec:topological-numbers}. We clarify that the bulk-edge correspondence is satisfied based on the obtained topological numbers, and discuss the contribution of edge states to dynamics. Section \ref{sec:discussion_conclusion} is devoted to discussion and conclusion.

\section{topological numbers of nonunitary Floquet systems with chiral symmetry}
\label{sec:topological-numbers}
We first derive two topological numbers $\nu_0$ and $\nu_\pi$ of nonunitary open Floquet systems, which highlight the bulk-edge correspondence in topological phases protected by chiral symmetry. 

\subsection{Biorthogonal basis and quasienergy}
\label{subsec:basis_energy}
A Floquet system is described by a time-evolution operator, which we write as $U$. Note that $U$ is generally nonunitary in the present work, which corresponds to an effective non-Hermitian Hamiltonian $H$ via the relation $U=\exp(-iH)$. The state after $t$ time steps is described by
\begin{equation}
\ket{\psi(t)}=U^t\ket{\psi(0)},
\label{eq:time-evolution}
\end{equation}
where $\ket{\psi(0)}$ is an initial state and $t$ is an integer. 
As we explain later, in order to define chiral symmetry and derive topological numbers, we need to define two time-evolution operators fitted in different time frames \cite{asboth2013bulk},
\begin{align}
U^\prime=AB,\ \ U^{\prime\prime}=BA,
\label{eq:time-evolution_operator}
\end{align}
where $A$ and $B$ are nonunitary operators. In a specific model which we treat in Secs. \ref{sec:model} and \ref{sec:bulk-edge_correspondence}, $A$ and $B$ are defined as in Eq. (\ref{eq:AB}). Since $U^\prime$ and $U^{\prime\prime}$ are related by a similarity transformation $A^{-1}U^\prime A=U^{\prime\prime}$, they share the same eigenvalues. Then, right eigenequations of $U^\prime$ and $U^{\prime\prime}$ are written as
\begin{align}
U^\prime\ket{\phi_\varepsilon^\prime}
=e^{-i\varepsilon}\ket{\phi_\varepsilon^\prime},\ 
U^{\prime\prime}\ket{\phi_\varepsilon^{\prime\prime}}
=e^{-i\varepsilon}\ket{\phi_\varepsilon^{\prime\prime}},
\label{eq:eigen-equation_right}
\end{align}
where $\varepsilon$ is the quasienergy. 
Note that quasienergies $\varepsilon$ need not be real since $A$ and $B$ are nonunitary.
We also introduce right eigenequations of $(U^\prime)^\dagger$ and $(U^{\prime\prime})^\dagger$
\begin{align}
(U^\prime)^\dagger\ket{\chi_\varepsilon^\prime}
=e^{i\varepsilon^\ast}\ket{\chi_\varepsilon^\prime},\ 
(U^{\prime\prime})^\dagger\ket{\chi_\varepsilon^{\prime\prime}}
=e^{i\varepsilon^\ast}\ket{\chi_\varepsilon^{\prime\prime}},
\label{eq:eigen-equation_left}
\end{align}
which are equivalent to Hermitian conjugations of left eigenequations of $U^\prime$ and $U^{\prime\prime}$, respectively.  We assume that eigenstates satisfy the biorthogonal normalization condition
\begin{align}
\naiseki{\phi_\varepsilon^\prime}{\chi_{\tilde{\varepsilon}}^\prime}=
\naiseki{\phi_\varepsilon^{\prime\prime}}
{\chi_{\tilde{\varepsilon}}^{\prime\prime}}=\delta_{\varepsilon\tilde{\varepsilon}}.
\label{eq:normalization}
\end{align}
While $\ket{\phi_\varepsilon^\prime}=\ket{\chi_\varepsilon^\prime}$ and $\ket{\phi_\varepsilon^{\prime\prime}}=\ket{\chi_\varepsilon^{\prime\prime}}$ when time-evolution operators are unitary, right eigenstates and left eigenstates are not related by the Hermitian conjugation when time-evolution operators are nonunitary. \\\indent

\subsection{Chiral symmetry}
\label{subsec:chiral-symmetry}
We define chiral symmetry of open Floquet systems through a constraint on $A$ and $B$ which are components of $U^\prime$ and $U^{\prime\prime}$;
\begin{align}
\Gamma B \Gamma^{-1}=A^\dagger,
\label{eq:chiral-symmetry_each}
\end{align}
where $\Gamma$ is a unitary operator which satisfies $\Gamma^2=1$. Equation (\ref{eq:chiral-symmetry_each}) is a sufficient condition that the time-evolution operators satisfy
\begin{align}
\Gamma U^\prime \Gamma^{-1}=(U^\prime)^\dagger,\ 
\Gamma U^{\prime\prime} \Gamma^{-1}=(U^{\prime\prime})^\dagger.
\label{eq:chiral-symmetry_total}
\end{align}
Equation (\ref{eq:chiral-symmetry_total}) is consistent with chiral symmetry of non-Hermitian Hamiltonians $\Gamma H \Gamma^{-1}=-H^\dagger$ where $U=\exp(-iH)$ in Floquet systems. Chiral symmetry in Eq.\ (\ref{eq:chiral-symmetry_each}), which is a more strict condition in comparison to Eq. (\ref{eq:chiral-symmetry_total}), plays a crucial role for deriving topological numbers, as we show in the following.
Note that chiral symmetry of non-Hermitian Hamiltonians is distinct from sublattice symmetry which transforms $H$ to $-H$, due to $H \neq H^\dagger$ \cite{kawabata2019symmetry}. 
From Eqs. (\ref{eq:eigen-equation_right}), (\ref{eq:eigen-equation_left}), and (\ref{eq:chiral-symmetry_each}), we can understand that $\Gamma\ket{\phi_\varepsilon^\prime}$ and $\Gamma\ket{\phi_\varepsilon^{\prime\prime}}$ are proportional to $\ket{\chi_{-\varepsilon^\ast}^\prime}$ and $\ket{\chi_{-\varepsilon^\ast}^{\prime\prime}}$,
\begin{align*}
\Gamma\ket{\phi_\varepsilon^\prime}
=\gamma_\varepsilon^\prime\ket{\chi_{-\varepsilon^\ast}^\prime},\ 
\Gamma\ket{\phi_\varepsilon^{\prime\prime}}
=\gamma_\varepsilon^{\prime\prime}
\ket{\chi_{-\varepsilon^\ast}^{\prime\prime}},
\end{align*}
respectively, where $\gamma_\varepsilon^\prime$ and $\gamma_\varepsilon^{\prime\prime}$ are proportionality coefficients. In one-dimensional Floquet systems with chiral symmetry, quasienergies of topologically protected edge states reside in a real line gap Re$(\varepsilon)=0$ or Re$(\varepsilon)=\pi$, since these lines are symmetric axes of quasienergy spectra. Therefore, we consider eigenstates whose quasienergies satisfy Re$(\varepsilon)=0$ or Re$(\varepsilon)=\pi$, which results in
\begin{align}
\Gamma\ket{\phi_\varepsilon^\prime}
=\gamma_\varepsilon^\prime\ket{\chi_\varepsilon^\prime},\ 
\Gamma\ket{\phi_\varepsilon^{\prime\prime}}
=\gamma_\varepsilon^{\prime\prime}
\ket{\chi_\varepsilon^{\prime\prime}},\ \text{Re}(\varepsilon)=0,\pi.
\label{eq:condition_chiral_zp}
\end{align}
The proportionality constant $\gamma_\varepsilon^\prime$ takes real values, since eigenstates satisfy normalization conditions in Eq. (\ref{eq:normalization}),
\begin{align*}
\naiseki{\chi_\varepsilon^\prime}
{\phi_\varepsilon^\prime}=\bra{\chi_\varepsilon^\prime}
\Gamma^2\ket{\phi_\varepsilon^\prime}
=\frac{\gamma_\varepsilon^\prime}{(\gamma_\varepsilon^\prime)^\ast}
\naiseki{\phi_\varepsilon^\prime}{\chi_\varepsilon^\prime}=1.
\end{align*}
In the same way, we can also find that $\gamma_\varepsilon^{\prime\prime}$ is real. Therefore, we can recognize the signs of $\gamma_\varepsilon^\prime$ and $\gamma_\varepsilon^{\prime\prime}$ as the labels of edge states, and we refer to the sign of $\gamma_\varepsilon'$ or $\gamma_\varepsilon''$ as chirality of each edge state.\\\indent

\subsection{Winding numbers}
\label{subsec:winding-number}
Winding numbers $\nu^\prime$ and $\nu^{\prime\prime}$ respectively for $U^\prime$ and $U^{\prime\prime}$ can be calculated by using the method proposed in \cite{esaki2011edge}. Following the proposed method, we consider situations in which the chiral symmetry operator is $\sigma_3$. For simplicity, we address $2\times2$ time-evolution operators in momentum space and focus only on $\tilde{U}'(k)$ which is obtained by Fourier transformation of $U'$. The right and left eigenequations of $\tilde{U}'(k)$ with eigenvalues $\lambda_\pm(k)$ are
\begin{align}
\tilde{U}'(k)\ket{\phi_\pm(k)}&=\lambda_\pm(k)\ket{\phi_\pm(k)},
\label{eq:right_eigenstate_general}\\
\bra{\chi_\pm(k)}\tilde{U}'(k)&=\bra{\chi_\pm(k)}\lambda_\pm(k).
\label{eq:left_eigenstate_general}
\end{align}
We assume $\langle\chi_\pm(k)|\phi_\mp(k)\rangle=0$ and $\langle\chi_\pm(k)|\phi_\pm(k)\rangle=1$. From these eigenstates, we define a Hermitian matrix $\tilde{Q}(k)$ as
\begin{align}
2\tilde{Q}(k)=\sum_{\sigma}\sigma\left(\ket{\phi_\sigma(k)}\bra{\chi_\sigma(k)}+\ket{\chi_\sigma(k)}\bra{\phi_\sigma(k)}\right),
\label{eq:Q_k}
\end{align}
where $\sigma=\pm$. Since chiral symmetry ensures $\sigma_3\ket{\phi_\pm(k)}=\gamma_\pm\ket{\chi_\mp(k)}$ where we can make the proportionality constant satisfy $|\gamma_\pm|^2=1$, $\{\tilde{Q}(k),\sigma_3\}=0$ is satisfied. Then, $\tilde{Q}(k)$ can be written as
\begin{align}
\tilde{Q}(k)
=\left(\begin{array}{cc}
0&q(k)\\
q^\ast(k)&0
\end{array}\right).
\label{eq:Q_k_off-diagonal}
\end{align}
From $\tilde{Q}(k)$, the winding number of $U^\prime$ is defined as
\begin{align}
\nu^\prime=\frac{1}{2\pi i}\oint dk q^{-1}(k)\frac{d}{dk}q(k).
\label{eq:nu_prime_definition}
\end{align}
From eigenstates of $\tilde{U}''(k)$, $\nu''$ can be obtained in the same way as $\nu'$. We will explain the method for a specific model later.

\subsection{Bulk-edge correspondence}
\label{subsec:bulk-edge_correspondence}
We relate the winding numbers $\nu^\prime$ and $\nu^{\prime\prime}$ with the number of edge states which satisfy Re$(\varepsilon)=0$ or $\pi$. To this end we consider that a system is composed of two adjacent regions labeled by L and R, and the time-evolution operators in the region L (R) have winding numbers $\nu_\text{L}^\prime$ and $\nu_\text{L}^{\prime\prime}$ ($\nu_\text{R}^\prime$ and $\nu_\text{R}^{\prime\prime}$). Then, we assume that the winding numbers satisfy
\begin{align}
\nu_\text{L}^\prime-\nu_\text{R}^\prime
=n_{0+}^\prime+n_{\pi+}^\prime
-n_{0-}^\prime-n_{\pi-}^\prime,
\label{eq:request_p}\\
\nu_\text{L}^{\prime\prime}-\nu_\text{R}^{\prime\prime}
=n_{0+}^{\prime\prime}+n_{\pi+}^{\prime\prime}
-n_{0-}^{\prime\prime}-n_{\pi-}^{\prime\prime},
\label{eq:request_pp}
\end{align}
where $n_{\text{Re}(\varepsilon)\pm}^\prime$ and $n_{\text{Re}(\varepsilon)\pm}^{\prime\prime}$ represent the numbers of edge states of $U^\prime$ and $U^{\prime\prime}$, whose subscripts denote the real part of $\varepsilon$ and chirality $\text{sign}(\gamma_\varepsilon^\prime),\,\text{sign}(\gamma_\varepsilon^{\prime\prime})$. The meaning of Eqs. (\ref{eq:request_p}) and (\ref{eq:request_pp}) is as follows. When edge states are shifted from Re$(\varepsilon)=0$ or $\pi$ due to a perturbation not breaking chiral symmetry, two states with different chirality must be paired \cite{esaki2011edge}. In this case, the number of edge states which remain on Re($\varepsilon)=0\ (\pi)$ becomes $n_{0+}^\prime-n_{0-}^\prime\ (n_{\pi+}^\prime-n_{\pi-}^\prime)$ for $U'$. This is the same for $U''$. Therefore, Eqs. (\ref{eq:request_p}) and (\ref{eq:request_pp}) are relations of the bulk-edge correspondence, which is confirmed in some situations \cite{esaki2011edge}. To satisfy Eqs. (\ref{eq:request_p}) and (\ref{eq:request_pp}), the winding numbers $\nu^\prime$ and $\nu^{\prime\prime}$ are either those calculated in homogeneous systems with periodic boundary conditions without non-Hermitian skin effect or those calculated using generalized Brillouin zone when the skin effect occurs \cite{yao2018edge,
kunst2018biorthogonal,
lee2019anatomy,
yokomizo2019non}. In this work, we use the former winding numbers in Sec. \ref{sec:bulk-edge_correspondence}. To our knowledge, it is empirically known that Eq. (\ref{eq:request_p}) is satisfied in chiral symmetric non-Hermitian systems when winding numbers calculated in homogeneous systems are used \cite{esaki2011edge}, where Eqs. (\ref{eq:request_p}) and (\ref{eq:request_pp}) are collapsed into a single equation with $n_{\pi\pm}=0$. Topological numbers which we want to derive are $\nu_0$ and $\nu_\pi$ satisfying
\begin{align}
\nu_0^\text{L}-\nu_0^\text{R}=n_{0+}^\prime-n_{0-}^\prime,\ 
\nu_\pi^\text{L}-\nu_\pi^\text{R}=n_{\pi+}^\prime-n_{\pi-}^\prime,
\label{eq:request_nu}
\end{align}
which predict the numbers of topologically protected edge states with Re$(\varepsilon)=0$ and Re$(\varepsilon)=\pi$, respectively.
\\\indent
We derive $\nu_0$ and $\nu_\pi$ by clarifying a relation between $\gamma_\varepsilon^\prime$ and $\gamma_\varepsilon^{\prime\prime}$. To this end, we operate $B$ and $A^\dagger$ on the right eigenequations of $U^\prime$ and $(U^\prime)^\dagger$ respectively, which results in
\begin{align}
BAB\ket{\phi_\varepsilon^\prime}
&=e^{-i\varepsilon}B\ket{\phi_\varepsilon^\prime},\ 
\label{eq:operation_right}\\
A^\dagger B^\dagger A^\dagger\ket{\chi_\varepsilon^\prime}
&=e^{i\varepsilon^\ast}A^\dagger\ket{\chi_\varepsilon^\prime}.
\label{eq:operation_left}
\end{align}
Then, from Eqs. (\ref{eq:time-evolution_operator})-(\ref{eq:eigen-equation_left}), (\ref{eq:operation_right}) and (\ref{eq:operation_left}), we can see that
\begin{align}
B\ket{\phi_\varepsilon^\prime}
=b\ket{\phi_\varepsilon^{\prime\prime}},\ 
A^\dagger\ket{\chi_\varepsilon^\prime}
=a\ket{\chi_\varepsilon^{\prime\prime}}
\label{eq:condition_operation}
\end{align}
are satisfied. The complex numbers $a$ and $b$ in Eq. (\ref{eq:condition_operation}) are related by
\begin{align}
a=\frac{e^{i\varepsilon^\ast}}{b^\ast},
\label{eq:condition_normalization_ab}
\end{align}
which is obtained by using Eq. (\ref{eq:condition_operation}) and the normalization conditions in Eq. (\ref{eq:normalization}), 
\begin{align*}
\naiseki{\phi_\varepsilon^{\prime\prime}}
{\chi_\varepsilon^{\prime\prime}}=\bra{\phi_\varepsilon^\prime}
\frac{1}{b^\ast}B^\dagger A^\dagger\frac{1}{a}
\ket{\chi_\varepsilon^\prime}
=\frac{e^{i\varepsilon^\ast}}{ab^\ast}
\naiseki{\phi_\varepsilon^\prime}{\chi_\varepsilon^\prime}=1.
\end{align*}
From Eqs. (\ref{eq:chiral-symmetry_each}), (\ref{eq:condition_chiral_zp}), and (\ref{eq:condition_operation}), $\Gamma\ket{\phi_\varepsilon^{\prime\prime}}$ becomes
\begin{align}
\Gamma\ket{\phi_\varepsilon^{\prime\prime}}
=\frac{1}{b}\Gamma B \Gamma^2\ket{\phi_\varepsilon^\prime}
=\frac{\gamma_\varepsilon^\prime}{b}A^\dagger
\ket{\chi_\varepsilon^\prime}=\frac{a}{b}
\gamma_\varepsilon^\prime\ket{\chi_\varepsilon^{\prime\prime}}.
\label{eq:transform_chirality}
\end{align}
Equations (\ref{eq:condition_normalization_ab}) and (\ref{eq:transform_chirality}) mean that $\gamma_\varepsilon^\prime$ and $\gamma_\varepsilon^{\prime\prime}$ satisfy
\begin{align}
\gamma_\varepsilon^{\prime\prime}=\frac{e^{i\varepsilon^\ast}}{|b|^2}
\gamma_\varepsilon^\prime.
\label{eq:relation_gamma}
\end{align}
From Eq. (\ref{eq:relation_gamma}), we can understand that edge states with Re$(\varepsilon)=0$ ($\pi$) have the same (opposite) chirality in systems described by $U^\prime$ and $U^{\prime\prime}$. This means that the numbers of edge states satisfy
\begin{align}
n_{0+}^\prime=n_{0+}^{\prime\prime},\ 
n_{\pi+}^\prime=n_{\pi-}^{\prime\prime},\ 
n_{0-}^\prime=n_{0-}^{\prime\prime},\ 
n_{\pi-}^\prime=n_{\pi+}^{\prime\prime}.
\label{eq:relation_number}
\end{align}
From Eqs. (\ref{eq:request_p})-(\ref{eq:request_nu}), and (\ref{eq:relation_number}), we can obtain
\begin{align}
\nu_0=\frac{\nu^\prime+\nu^{\prime\prime}}{2},\ \ 
\nu_\pi=\frac{\nu^\prime-\nu^{\prime\prime}}{2}.
\label{eq:nu}
\end{align}
Equation (\ref{eq:nu}) has been employed in nonunitary Floquet systems \cite{xiao2017observation,
xiao2018higher,
chen2018characterization,
kawasaki2020bulk} based on the analogy to unitary Floquet systems in which the same formula is proven to be satisfied \cite{asboth2013bulk}. Our derivation gives the microscopic foundation of Eq. (\ref{eq:nu}) that has been used empirically so far for nonunitary open Floquet systems.

\section{Model}
\label{sec:model}
\begin{figure}
\begin{center}
\includegraphics[width=9cm]{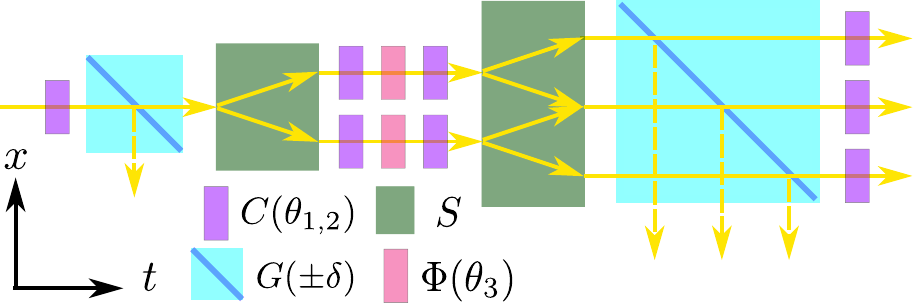}
\caption{(Color online) A schematic picture for one time step in the case of photonic quantum walks governed by $U'$ in Eq. (\ref{eq:U}), where photons pass through arrayed optical elements. The direction indicated by yellow arrows in which photons proceed is considered to be the direction of time. In experiments \cite{xiao2017observation,xiao2018higher,xiao2020non}, combinations of wave plates correspond to coin operators $C(\theta_{1,2})$ and the phase operator $\Phi(\theta_3)$. The shift operator $S$ and gain-loss operators $G(\pm\delta)$ respectively describe effects of beam displacers and partially polarizing beam splitters. }
\label{fig:experimental-setting}
\end{center}
\end{figure}

We introduce the time-evolution operator of the one-dimensional nonunitary quantum walk as an example of the chiral symmetric open Floquet systems. Our model is similar to the nonunitary quantum walk realized in the experiment in Ref.\ \cite{regensburger2012parity} and studied further in Ref.\ \cite{mochizuki2016explicit}, though chiral symmetry in Eq. \ref{eq:chiral-symmetry_each} was not mentioned in the previous works. The basis of the walker is defined from one-dimensional position space $\ket{x}$ and two internal states $\ket{{\cal L}}:=(1,0)^\text{T}$ and $\ket{{\cal R}}:=(0,1)^\text{T}$ where $x\in \mathbb{Z}$ and the superscript T is the transpose. Thereby, a state at a time step $t$ is written as 
\begin{align}
\ket{\psi(t)}=\sum_{x \in \mathbb{Z}}\sum_{s={\cal L,R}}\psi_{x,s}(t)\ket{x}\otimes\ket{s}.
\label{eq:wave-function}
\end{align}
The time-evolution operator $U^\prime$ for one time step of the one-dimensional nonunitary quantum walk is defined by the product of elemental operators,
\begin{align}
U^\prime:=C(\theta_1)SG(\delta)C(\theta_2)\Phi(\theta_3)C(\theta_2)SG(-\delta)C(\theta_1),
\label{eq:U}
\end{align}
where each elemental operator is defined as
\begin{align*}
&C(\theta_{j=1,2}):=\sum_x \ket{x}\bra{x}\otimes\tilde{C}[\theta_j(x)],\\
&\quad\tilde{C}[\theta_j(x)]:=\left(\begin{array}{cc}
\cos[\theta_j(x)/2]&i\sin[\theta_j(x)/2]\\
i\sin[\theta_j(x)/2]&\cos[\theta_j(x)/2]
\end{array}\right)=e^{i\frac{\theta_j(x)}{2}\sigma_1},\nonumber\\
&S:=\sum_x\ket{x-1}\bra{x}\otimes\ket{{\cal L}}\bra{{\cal L}}
+\ket{x+1}\bra{x}\otimes\ket{{\cal R}}\bra{{\cal R}}\\
&\quad=\sum_k\ket{k}\bra{k}\otimes\tilde{S}(k),\ 
\tilde{S}(k):=\left(\begin{array}{cc}
e^{ik} & 0 \\
0 & e^{-ik}
\end{array}\right)=e^{ik\sigma_3},\\
&G(\delta):=\sum_x\ket{x}\bra{x}\otimes\tilde{G}[\delta(x)],\\
&\quad \tilde{G}[\delta(x)]:=\left(\begin{array}{cc}
e^{\delta(x)} & 0 \\
0 & e^{-\delta(x)}
\end{array}\right)=e^{\delta(x)\sigma_3},\\
&\Phi(\theta_3):=\sum_x\ket{x}\bra{x}\otimes\tilde{\Phi}[\theta_3(x)],\\ 
&\quad\tilde{\Phi}[\theta(x)]:=\left(\begin{array}{cc}
e^{i\theta_3(x)} & 0 \\
0 & e^{-i\theta_3(x)}
\end{array}\right)=e^{i\theta_3(x)\sigma_3}.
\end{align*}
Here, $\sigma_i$ $(i=1,2,3)$ are Pauli matrices. The coin operator $\tilde{C}[\theta_j(x)]$ changes the internal states of walkers through the position dependent $\theta_j(x)$ and the shift operator $S$ shifts a walker to its adjacent site depending on its internal state. The gain and loss operator $\tilde{G}[\delta(x)]$ with positive $\delta(x)$ amplifies (damps) the wavefunction amplitude with the internal state $\ket{{\cal L}}$ ($\ket{{\cal R}}$) by the factor $e^{\delta(x)}$ ($e^{-\delta(x)}$), and $\tilde{G}[-\delta(x)]$ vice versa. In a similar way, the phase operator $\tilde{\Phi}[\theta_3(x)]$ induces the phase $e^{i\theta_3(x)}$ ($e^{-i\theta_3(x)}$) to the wavefunction amplitude of the internal state $\ket{\cal L}$ ($\ket{\cal R}$) at the position $x$. The time evolution is described by $U^\prime$, following Eq. (\ref{eq:time-evolution}). Figure \ref{fig:experimental-setting} shows a schematic picture of $U'$ in Eq. (\ref{eq:U}) implemented by photonic quantum walks in Refs. \cite{xiao2017observation,
xiao2018higher,
xiao2020non}. We write the eigenvalue of $U^\prime$ as $\lambda$,
\begin{align}
U^\prime\ket{\phi_\lambda} = \lambda \ket{\phi_\lambda},\ \ \lambda=e^{-i\varepsilon},
\label{eq:quasienergy}
\end{align}
where $\varepsilon$ is the quasienergy. When $\delta \ne 0$, $U^\prime$ becomes nonunitary and $|\lambda|\ne 1$, which makes the quasienergy $\varepsilon$ complex, in general. 

\subsection{Symmetries}
\label{subsec:symmetries}
The nonunitary time-evolution operator $U^\prime$ in Eq.\ (\ref{eq:U}) has various symmetries, while some of symmetries exist only if parameters satisfy specific conditions.
We briefly summarize relevant symmetries which are important to argue topological phases and dynamics by considering constraints on the parameters. 

\subsubsection{In the case of no constraints}
\label{subsubsec:AIII}
First, we assume no constraints on the parameters of the time-evolution operator $U^\prime$. Therefore, values of $\theta_j(x)$ and $\delta(x)$ become arbitrary.
In this case, the time-evolution operator in Eq.\ (\ref{eq:U}) retains only chiral symmetry in Eq. (\ref{eq:chiral-symmetry_each}).  By decomposing the time-evolution operator into $U^\prime=AB$ where $A$ and $B$ are defined as 
\begin{subequations}
\begin{align}
A&=C(\theta_1)G(\delta)SC(\theta_2)\Phi(\theta_3/2),
\label{eq:A}\\
B&=\Phi(\theta_3/2)C(\theta_2)SG(-\delta)C(\theta_1),
\label{eq:B}
\end{align}
\label{eq:AB}
\end{subequations}
chiral symmetry is confirmed with
\begin{align}
\Gamma=\sum_x\ket{x}\bra{x}\otimes\sigma_2.
\label{eq:Gamma}
\end{align}
Therefore, the time-evolution operator without any constraint belongs to class AIII.
It is known that class AIII can have nontrivial $\mathbb{Z}$ topological phases in one dimension \cite{kawabata2019symmetry}. We remark that nonunitary quantum walks which belong to class AIII have not been studied so far. 

\subsubsection{In the case of $\theta_3(x)=0,\, \pi$ }
\label{subsubsec:BDI_dagger}
Next, we consider the case of $\theta_3(x)=0$ or $\pi$, but no constraints on $\theta_{1/2}(x)$ and $\delta(x)$. In this case, in addition to chiral symmetry, $U^\prime$ satisfies particle-hole symmetry and time-reversal symmetry in AZ$^\dagger$ classification,
\begin{align}
\mathcal{C}U^\prime\mathcal{C}^{-1}&=U^\prime,
\label{eq:particle-hole_symmetry}\\
T U^\prime T^{-1}&=(U^\prime)^\dagger,
\label{eq:time-reversal_symmerty}
\end{align}
respectively, where the symmetry operators are
\begin{align}
\mathcal{C}&=\sum_x\ket{x}\bra{x}\otimes\sigma_3\mathcal{K},
\label{eq:C}\\
T&=\sum_x\ket{x}\bra{x}\otimes\sigma_1\mathcal{K}.
\label{eq:T}
\end{align}
Here $\mathcal{K}$ is the complex conjugation operator. 
Particle-hole symmetry and time-reversal symmetry in Eqs. (\ref{eq:particle-hole_symmetry}) and (\ref{eq:time-reversal_symmerty}) are defined from these symmetries of non-Hermitian Hamiltonians $\mathcal{C}H\mathcal{C}^{-1}=-H$ and $THT^{-1}=-H^\dagger$, respectively, using $U=\exp(-iH)$. Then, the system is classified into class BDI$^\dagger$ in the AZ$^\dagger$ classification and possibly possesses nontrivial topological phases \cite{kawabata2019symmetry}. When $\delta(x)=0$, the time-evolution operator is unitary $(U^\prime)^\dagger=(U^\prime)^{-1}$ and classified into class BDI in the AZ classification \cite{schnyder2008classification}. Note that Refs. \cite{mochizuki2016explicit,mochizuki2017effects,xiao2017observation} which explore similar quantum walks to our model do not mention chiral, particle-hole, and time-reversal symmetries in Eqs. (\ref{eq:chiral-symmetry_each}), (\ref{eq:particle-hole_symmetry}), and (\ref{eq:time-reversal_symmerty}). We reveal these symmetries for the first time.\\\indent

\subsubsection{In the case of $\theta_j(x)=\theta_j(-x)$ and $\delta(x)=\delta(-x)$}
\label{subsubsec:PT-symmetric}
One of the interesting symmetries for open systems is $\mathcal{PT}$ symmetry, since the presence of this symmetry can result in real quasienergies.
The nonunitary time-evolution operator in Eq.\ (\ref{eq:U}) also retains $\mathcal{PT}$ symmetry if parameters $\theta_j(x)$ and $\delta(x)$ satisfy a specific position dependence.
$\mathcal{PT}$ symmetry in open Floquet systems requires the nonunitary time-evolution operator to satisfy \cite{mochizuki2016explicit,mochizuki2017effects}
\begin{align}
(\mathcal{PT})U^\prime(\mathcal{PT})^{-1}=(U^\prime)^{-1}.
\label{eq:PT_symmetry}
\end{align}
Equation (\ref{eq:PT_symmetry}) is satisfied by defining $\mathcal{PT}$ symmetry operator as 
\begin{align}
\mathcal{PT}=\sum_x\ket{-x}\bra{x}\otimes\sigma_0\mathcal{K},
\label{eq:PT}
\end{align}
with constraints on position dependences of parameters,
\begin{align}
\theta_j(x)&=\theta_j(-x),\quad
\delta(x)=\delta(-x),
\label{eq:condition_PT}
\end{align}
where $j=1,2,3$. When the time-evolution operator possesses $\mathcal{PT}$ symmetry and the eigenstate $\ket{\phi_\lambda}$ is also an eigenstate of $\mathcal{PT}$ symmetry operator, $|\lambda|=1$ and $\varepsilon$ becomes real \cite{bender2007making}.
As we clarify in Sec. \ref{subsec:dynamics}, $\mathcal{PT}$ symmetry plays a crucial role in the dynamics of the nonunitary quantum walk.

\subsection{Spectral properties in a homogeneous system}
\begin{figure*}
\begin{center}
\includegraphics[width=17cm]{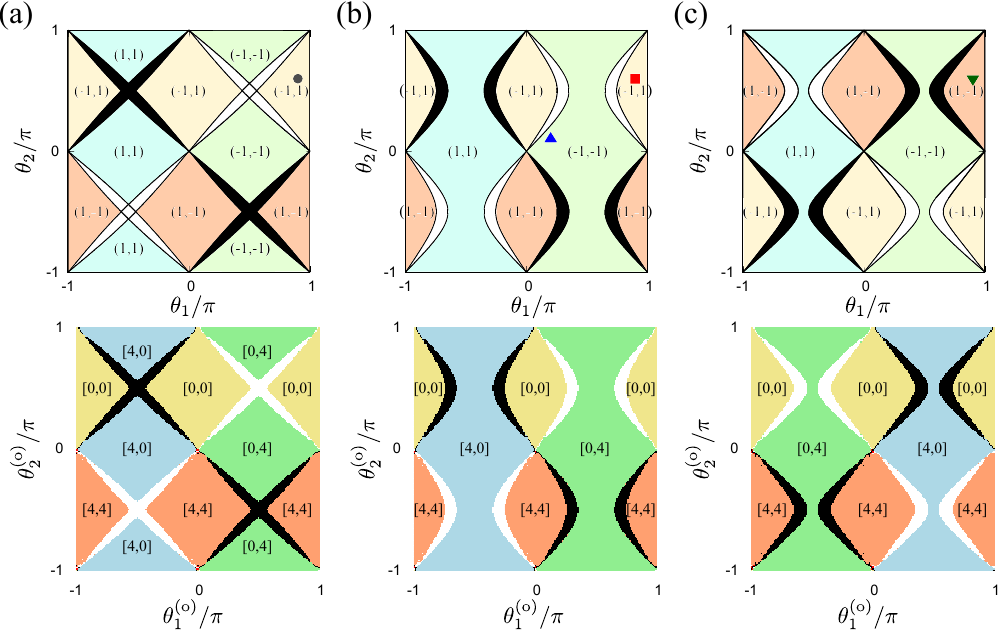}
\caption{(Color online) The top row: The phase diagrams of topological numbers $(\nu_0,\nu_\pi)$ as functions of $\theta_1$ and $\theta_2$ of the nonunitary quantum walk in Eq.\ (\ref{eq:U}) with $e^\delta=1.1$ and (a) $\theta_3=0$, (b) $\theta_3=\pi/5$, (c) $\theta_3=9\pi/10$. The system for (a) is classified in class BDI$^\dagger$, the systems for (b) and (c) correspond to class AIII. The regions with topological numbers $(\nu_0,\nu_\pi)$ correspond to unbroken $\mathcal{PT}$ symmetry phases, while black and white regions represent broken $\mathcal{PT}$ symmetry phases with complex quasienergies whose real parts are $\text{Re}(\varepsilon)=0$ and $\pi$, respectively. The gray circle, red rectangle, and green lower triangle respectively in (a), (b), and (c) represent $(\theta_1^\text{(i)},\theta_2^\text{(i)})=(9\pi/10, 3\pi/5)$, while the blue upper rectangle in (b) represents $(\theta_1^\text{(o)},\theta_2^\text{(o)})=(\pi/5,\pi/10)$ which is used in Fig.\ \ref{fig:eigen-value_state}. The bottom row: The numbers of edge states $[N_0,N_\pi]$ appearing at the quasienergy $\text{Re}(\varepsilon)=0,\pi$ for various values of $(\theta_1^\text{(o)},\theta_2^\text{(o)})$ in the system described in Fig. \ref{fig:system} and Eq. (\ref{eq:cnd:PT:theta}) when $(\theta_1^\text{(i)},\theta_2^\text{(i)})$  in Eq.\ (\ref{eq:theta_in}) are fixed at $(9\pi/10, 3\pi/5)$ and other parameters are the same as those in the top row.}
\label{fig:phase-diagram}
\end{center}
\end{figure*}

We study the spectral properties of the nonunitary quantum walk and clarify a condition for the presence of finite band gaps of the quasienergy spectrum around Re($\varepsilon)=0$ and $\pi$. When both gaps are open, topological numbers for the system with chiral symmetry are well defined. We assume a homogeneous system in which $\theta_j(x)=\theta_j$ and $\delta(x)=\delta$ are constant. Since the conditions in Eq.\ (\ref{eq:condition_PT}) are satisfied, $\mathcal{PT}$ symmetry is retained in the homogeneous system and the quasienergies become entirely real in certain parameter spaces. The time-evolution operator in momentum space is obtained by Fourier transformation,
\begin{align}
\tilde{U}^\prime(k)&=d_0(k)\sigma_0+id_1(k)\sigma_1+d_2(k)\sigma_2+id_3(k)\sigma_3,
\label{eq:U_prime_k}\\
d_0(k)&=\cos\theta_3(\cos\theta_1\cos\theta_2\cos2k-\sin\theta_1\sin\theta_2\cosh2\delta)\nonumber\\
&-\cos\theta_1\sin\theta_3\sin2k,\nonumber\\
d_1(k)&=\cos\theta_3(\sin\theta_1\cos\theta_2\cos2k+\cos\theta_1\sin\theta_2\cosh2\delta)\nonumber\\
&-\sin\theta_1\sin\theta_3\sin2k,\nonumber\\
d_2(k)&=d_2=\sin\theta_2\cos\theta_3\sinh2\delta,\nonumber\\
d_3(k)&=\cos\theta_2\cos\theta_3\sin2k+\sin\theta_3\cos2k,\nonumber
\end{align}
where $d_0^2(k)+d_1^2(k)-d_2^2+d_3^2(k)=1$ is satisfied. Then, the eigenvalues of $\tilde{U}^\prime(k)$ are derived as 
\begin{align}
\lambda_{\pm}(k)=d_0(k)\pm i\sqrt{1-{d_0^2(k)}}.
\label{eq:eigenvalue}
\end{align}
On one hand, when $|d_0(k)| \le 1$ for any $k\in[0,2\pi)$, all of the quasienergies are kept real since $|\lambda_\pm(k)|=1$, although $U^\prime$ is nonunitary. On the other hand, when $|d_0(k)| > 1$ in a certain range of $k$, $\lambda_\pm(k)=d_0(k)\mp \sqrt{d_0^2(k)-1}\ (\ne \pm1)$, and then the corresponding quasienergies become complex in the range. The former and latter situations are called an unbroken $\mathcal{PT}$ symmetry phase and a broken $\mathcal{PT}$ symmetry phase, respectively \cite{bender2007making}. The condition for the presence or absence of the quasienergy band gap around Re($\varepsilon)=0$ or $\pi$, is also discerned by the above unbroken/broken $\mathcal{PT}$ symmetry phases. 
To derive the condition of the finite band gaps, we rewrite $d_0(k)$ as
\begin{align}
 d_0(k) &= \alpha\cos(2k+\beta)-\sin\theta_1\sin \theta_2\cos\theta_3\cosh2\delta,\nonumber\\
 \alpha&=\sqrt{\cos^2\theta_1\cos^2\theta_2\cos^2\theta_3+\cos^2\theta_1\sin^2\theta_3},\nonumber\\
 \cos\beta&=\sin\theta_1\sin\theta_2\cos\theta_3/\alpha,\ \sin\beta=\cos\theta_1\sin\theta_3/\alpha.\nonumber
\end{align}
Since the band gaps are closed at $\lambda=d_0(k_*)=\pm 1$ at a specific $k_*$, 
the condition of the finite band gaps is derived as
\begin{align}
\frac{|\sin\theta_1\sin\theta_2\cos\theta_3\cosh2\delta\pm1|}{\sqrt{\cos^2\theta_1\cos^2\theta_2\cos^2\theta_3+ \cos^2\theta_1\sin^2\theta_3 }}
>1,
\label{eq:band_gap}
\end{align}
where $\pm1$ in the numerator of the left hand side corresponds to the condition of the finite band gaps at $\lambda=\pm1$. In the following section, we use Eq.\ (\ref{eq:band_gap}) to draw the phase diagram of topological numbers in Fig. \ref{fig:phase-diagram}.

\section{Bulk-Edge correspondence}
\label{sec:bulk-edge_correspondence}
In this section, we derive the topological numbers from the time-evolution operators in Eq. (\ref{eq:time-evolution_operator}) with Eq.\ (\ref{eq:AB}) and demonstrate the bulk-edge correspondence in two different symmetry classes BDI$^\dagger$ and AIII.
\subsection{Topological numbers}
\label{subsec:topological-numbers}
Based on chiral symmetry of $U^\prime$, we calculate topological numbers $(\nu_0,\nu_\pi)$ of the system. Since $A$ and $B$ in Eq. (\ref{eq:AB}) satisfy Eq. (\ref{eq:chiral-symmetry_each}) with $\Gamma$ in Eq. (\ref{eq:Gamma}), we can use Eq. (\ref{eq:nu}) to derive $\nu_0$ and $\nu_\pi$. We note that the bulk-edge correspondence studied in closed Hermitian systems can be broken in open systems due to non-Hermitian skin effect in which bulk spectra drastically depend on boundary conditions. This occurs when spectra under periodic boundary conditions form a closed loop in the complex plane \cite{okuma2020topological,zhang2020correspondence}. However, in our model, spectra without boundaries do not form any closed loop, and bulk spectra never experience drastic changes originating from boundary conditions. Therefore, we calculate the winding number $\nu^\prime$ from eigenstates of homogeneous $U^\prime$ with periodic boundary conditions, {\it i.e.} eigenstates of $\tilde{U}^\prime(k)$, using a method proposed in Ref. \cite{esaki2011edge}. 
Note that $\mathcal{PT}$ symmetry exists in the argument below because the system is homogeneous.
The calculation of $\nu^{\prime\prime}$ from $U^{\prime\prime}=BA=\Phi(\theta_3/2)C(\theta_2)SG(-\delta)C(\theta_1)G(\delta)C(\theta_2)\Phi(\theta_3/2)$ is the same as that of $\nu^\prime$.\\\indent
In order to follow the procedure explained in Sec. \ref{subsec:winding-number}, we apply a unitary transformation to the time-evolution operator,
\begin{align}
\tilde{V}(k)=e^{i\frac{\pi}{4}\sigma_1}\tilde{U}^\prime(k)e^{-i\frac{\pi}{4}\sigma_1},
\label{eq:V}
\end{align}
which makes chiral symmetry operator become $\sigma_3$. When $\mathcal{PT}$ symmetry is preserved and the gaps around Re($\varepsilon)=0,\pi$ are open, the right and left eigenstates of $\tilde{V}(k)$ with eigenvalues $\lambda_\pm(k)$ are
\begin{align}
\ket{\phi_\pm(k)}=&\frac{1}{\sqrt{2\cos2\Omega_{k}}}(e^{\pm i\Omega_{k}},\pm ie^{\mp i\Omega_{k}}e^{-i\vartheta_{k}})^\text{T},
\label{eq:right_eigenstate_model}\\
\bra{\chi_\pm(k)}=&\frac{1}{\sqrt{2\cos2\Omega_{k}}}(e^{\pm i\Omega_{k}},\mp ie^{\mp i\Omega_{k}}e^{i\vartheta_{k}}),
\label{eq:left_eigenstate_model}
\end{align}
respectively, where $\vartheta_k$ and $\Omega_k$ are defined as
\begin{align}
|d(k)|e^{i\vartheta_k}=&d_3(k)+id_1(k)
\label{eq:definition1}\\
\cos 2\Omega_k=&\sqrt{1-\left(\frac{d_2}{|d(k)|}\right)^2},\ \ \sin 2\Omega_k=\frac{d_2}{|d(k)|}.
\label{eq:definition2}
\end{align}
Substituting Eqs. (\ref{eq:right_eigenstate_model}) and (\ref{eq:left_eigenstate_model}) into Eq. (\ref{eq:Q_k}), $q(k)$ in Eq. (\ref{eq:Q_k_off-diagonal}) becomes
\begin{align}
q(k)=-\frac{i e^{i\vartheta_k}}{\cos2\Omega_k},
\label{eq:q_k}
\end{align}
which results in
\begin{align}
\nu^\prime=\frac{1}{2\pi i}\oint dk\frac{d}{dk}\ln[q(k)]=\frac{1}{2\pi}\oint d\vartheta_k.
\label{eq:nu_prime}
\end{align}
Calculating $\nu^{\prime\prime}$ in the same way and substituting $\nu^\prime$ and $\nu^{\prime\prime}$ into Eq. (\ref{eq:nu}), we can obtain topological numbers of the system.

The results with $e^\delta=1.1$ are shown in the upper row of Fig.\ \ref{fig:phase-diagram} with various $\theta_3$. While the system belongs to class BDI$^\dagger$ in the case of $\theta_3=0$ [Fig.\ \ref{fig:phase-diagram} (a)], the system is classified into class AIII in the case of $\theta_3\neq0$ [Fig.\ \ref{fig:phase-diagram} (b) and (c)]. We note that the topological numbers for the nonunitary quantum walk defined in Eq.\ (\ref{eq:nu}) are the same as those for the unitary quantum walk $(\delta=0)$ as long as both band gaps around Re($\varepsilon)=0$ and $\pi$ are open. Therefore, the topological numbers are robust against effects of gain and loss in the chiral symmetric nonunitary Floquet system.  
\begin{figure}[b]
\begin{center}
\includegraphics[width=7.5cm]{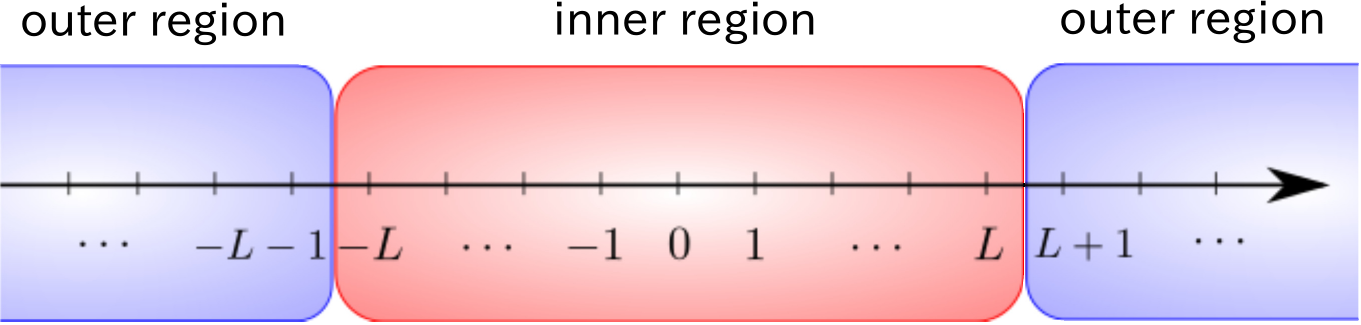}
\caption{(Color online) A schematic view of the $\mathcal{PT}$ symmetric nonunitary quantum walk defined in Eq.\ (\ref{eq:cnd:PT:theta}).}
\label{fig:system}
\end{center}
\end{figure}
For small $\delta$, increasing $\theta_3$ from $0$ to $\pi/5$ as shown in Fig.\ \ref{fig:phase-diagram} (a) and (b), the band gaps around Re($\varepsilon)=0$ and $\pi$ open for any values of $\theta_2$ around $\theta_1=\pm \pi/2$ and the regions with topological numbers $(\nu_0,\nu_\pi)=(\pm1,\pm1)$ are connected. The topological numbers $(\nu_0,\nu_\pi)=(\pm1,\mp1)$ are unchanged as long as the band gaps are open, while the regions are getting small as $\theta_3$ is increased from $0$ to $\pi/2$. At $\theta_3=\pi/2$, the regions with $(\nu_0,\nu_\pi)=(\pm1,\mp1)$ vanish. In this case, both band gaps around Re($\varepsilon)=0$ and $\pi$ are closed only when $\theta_1=0,\pi$ since Eq.\ (\ref{eq:band_gap}) with $\theta_3=\pi/2$ is satisfied excepting $\theta_1=0,\pi$. With increasing $\theta_3$ furthermore, the regions with $(\nu_0,\nu_\pi)=(\pm1,\mp1)$ appear again as shown in the top panel of Fig. \ref{fig:phase-diagram} (c), while signs of the topological numbers are different from those with $-\pi/2<\theta_3<\pi/2$ due to the band gap closing. 

\subsection{Counting the edge states}
\label{subsec:counting_edge-states}
Having established the phase diagrams in the top row of Fig. \ref{fig:phase-diagram}, we confirm that the topological numbers give the correct numbers of edge states even for the nonunitary Floquet topological phases with  chiral symmetry. In order to induce edge states, we spatially change topological numbers through position-dependent angles $\theta_j(x)$ of the coin operators taking the conditions in Eq.\ (\ref{eq:condition_PT}) into consideration. Accordingly, the system is separated into three, an inner and two outer regions as shown in Fig.\ \ref{fig:system}, which are discerned by the values of $\theta_1$ and $\theta_2$,
\begin{eqnarray}
\theta_j(x)\in
\begin{cases}
[\theta_j^\text{(o)}-w_j,\theta_j^\text{(o)}+w_j] & \left(x\le-L-1\right),\\
[\theta_j^\text{(i)}-w_j,\theta_j^\text{(i)}+w_j] & \left(-L\le x\le L\right),\\
[\theta_j^\text{(o)}-w_j,\theta_j^\text{(o)}+w_j] & \left(x\geq L+1\right),
\end{cases}
\label{eq:cnd:PT:theta}
\end{eqnarray} 
where $j=1,2,3$. The angles $\theta_j(x)$ are randomly distributed over the position space obeying box distributions whose widths are $w_j$ with the mean values $\theta_j^\text{(i)},\ \theta_j^\text{(o)}$. Also, we make $\delta(x)$ obey $\delta(x)\in[\delta-w_\delta,\delta+w_\delta]$, which is different from the position dependence of $\theta_j(x)$ because $\delta$ has no effects on $\nu_0$ and $\nu_\pi$ as long as the gaps are open. Since there are two interfaces near $x=\pm L$ at which the topological numbers vary, the time-evolution operator preserves $\mathcal{PT}$ symmetry when $w_j=w_\delta=0$, while nonzero $w_j,w_\delta$ results in the complex quasienergies due to the absence of $\mathcal{PT}$ symmetry. Note that the existence of two interfaces to induce edge states is in contrast to settings in unitary quantum walks \cite{kitagawa2012observation,chen2018observation} where $\mathcal{PT}$ symmetry is not taken into account and no spatial constraints on $\theta_j(x)$ are required. In the following numerical simulations, we fix the parameters as
\begin{equation}
e^\delta=1.1,\ \ 
(\theta^\text{(i)}_1,\,
\theta_2^\text{(i)})=
(9\pi/10,\,3\pi/5).
\label{eq:theta_in} 
\end{equation}
\begin{figure}[b]
\begin{center}
\includegraphics[width=8cm]{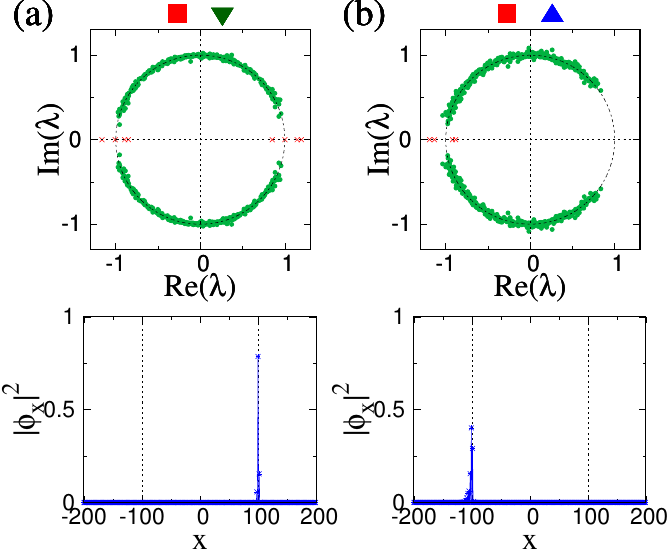}
\caption{(Color online) The eigenvalues and edge states of the time-evolution operator with chiral symmetry but without $\mathcal{PT}$ symmetry. We set $w_j=3\pi/20$ for all $j$ and $w_\delta=3/20$. (a) Parameters in the inner and outer regions respectively correspond to the red rectangle in Fig. \ref{fig:phase-diagram} (b) and the green lower triangle in Fig. \ref{fig:phase-diagram} (c); $\theta_1^\text{(o)}=\theta_1^\text{(i)}=9\pi/10$, $\theta_2^\text{(o)}=\theta_2^\text{(i)}=3\pi/5$, $\theta_3^\text{(i)}=\pi/5$, and $\theta_3^\text{(o)}=9\pi/10$. (b) Parameters in the inner and outer regions are respectively represented as the red rectangle and the blue upper triangle in Fig. \ref{fig:phase-diagram} (b); $\theta_1^\text{(o)}=\pi/5$, $\theta_2^\text{(o)}=\pi/10$, and $\theta_3^\text{(i)}=\theta_3^\text{(o)}=\pi/5$. In the top row, eigenvalues of topologically protected edge states are depicted as red crosses. In the bottom row, intensities of topological edge states are plotted, where dashed lines represent boundaries at which topological numbers change.}
\label{fig:eigen-value_state}
\end{center}
\end{figure}
When we calculate eigenvalues of the time-evolution operator $U'$ by numerical diagonalizations, the periodic boundary conditions are imposed on two ends of the finite system with $-M\le x\le M-1$, $M=2L$. In Fig.\ \ref{fig:eigen-value_state}, we show the eigenvalues and edge states of the time-evolution operator for the system in Eq.\ (\ref{eq:cnd:PT:theta}) with $\theta_{1,2}^\text{(i)}$ in Eq.\ (\ref{eq:theta_in}), $w_j/\pi=w_\delta=3/20$, and two sets of $\theta_{1,2,3}^\text{(o)}$ and $\theta_3^\text{(i)}$. Figure \ref{fig:eigen-value_state} (a) shows the results for $(\theta_1^\text{(o)},\theta_2^\text{(o)},\theta_3^\text{(i)},\theta_3^\text{(o)})=(9\pi/10,3\pi/5,\pi/5, 9\pi/10)$ with Eq.\ (\ref{eq:theta_in}), corresponding to the red rectangle in Fig.\ \ref{fig:phase-diagram} (b) and the green lower triangle in Fig.\ \ref{fig:phase-diagram} (c) respectively in the inner and outer regions. We expect the existence of two edge states on Re$(\varepsilon)=0$ and $\pi$ near each interface from the bulk-edge correspondence. In the same way, two edge states on Re($\varepsilon)=\pi$ near each interface are expected to appear in the case of Fig.\ \ref{fig:eigen-value_state} (b) where parameters $(\theta_1^\text{(o)},\theta_2^\text{(o)},\theta_3^\text{(i)},\theta_3^\text{(o)})=(\pi/5,\pi/10,\pi/5,\pi/5)$ with Eq. (\ref{eq:theta_in}) correspond to the red rectangle and the blue upper triangle in Fig. \ref{fig:phase-diagram} (b). The results of numerical diagonalizations in the top row of Fig.\ \ref{fig:eigen-value_state} clearly confirm the expectations since eigenvalues of topologically protected edge states (red crosses) satisfy Re($\varepsilon)=0$ or Re($\varepsilon)=\pi$, which is consistent with the discussion in Sec. \ref{sec:topological-numbers}. Among four eigenstates corresponding to eigenvalues on Re($\varepsilon)=0$ or $\pi$, two states are localized at the left boundary and the other two states reside in the right boundary, of which some states are shown in the bottom row of Fig. \ref{fig:eigen-value_state}. Thereby the bulk-edge correspondence is satisfied in the chiral symmetric open Floquet system. The eigenvalues which correspond to bulk states (green circles in Fig.\ \ref{fig:eigen-value_state}) fluctuate around the unit circle since the randomness of $\theta_j(x)$ and $\delta(x)$ breaks $\mathcal{PT}$ symmetry of the time-evolution operator. For the same reason, there are no pair of edge states having the opposite sign of the imaginary part of quasienergy, $\pm \text{Im}(\varepsilon)$, while such a pair appears for the chiral symmetric nonunitary Floquet system with $\mathcal{PT}$ symmetry.

Further we systematically check the above results for the whole parameter regions of $(\theta_1^\text{(o)},\theta_2^\text{(o)})$  in the following way. We focus on $\theta_3^\text{(i)}=\theta_3^{\text{(o)}}$ whose value is the same with each case of the top row in Fig.\ \ref{fig:phase-diagram}. Therefore, the parameters for the inner region are fixed. We count the numbers of eigenstates with $\text{Re}(\varepsilon)=0,\pi$ for various $\theta_{1,2}^\text{(o)}$ by treating two different system sizes $M=100$ and $M=200$. Then, we distinguish edge states from other states on the real axis due to the closing of band gaps on Re($\varepsilon)=0$ or $\pi$ by the system size dependence of the numbers. If the numbers of eigenstates with $\text{Re}(\varepsilon)=0$ and $\pi$ are unchanged with changing the system size $M$, the numbers are recognized as the numbers of edge states $N_0$ and $N_\pi$, respectively, originating from Floquet topological phases. In order to explicitly compare analytical results of clean systems shown in the top row of Fig. \ref{fig:phase-diagram} with numerical results obtained from the procedure explained above, we set $w_j=w_\delta=0$ in the present analysis. The results are summarized in the bottom row of Fig.\ \ref{fig:phase-diagram} showing the set $[N_0,N_\pi]$. Comparing the top and bottom rows of Fig.\ \ref{fig:phase-diagram} in the light of Eq.\ (\ref{eq:theta_in}), in unbroken $\mathcal{PT}$ symmetry phases, the numbers of edge states completely agree with the predictions by the bulk-edge correspondence. In broken $\mathcal{PT}$ symmetry phases, we observe that the numbers of eigenstates with $\text{Re}(\varepsilon)=0$ and/or $\pi$ increase as the size $M$ is increased, except a few points where the numbers show no system size dependence. However, this would be due to finite size effects and could be improved by using larger $M$. Thereby, we conclude that the bulk-edge correspondence works even for topological phases in chiral symmetric open Floquet systems classified into class BDI$^\dagger$ and AIII. We remark that Fig. \ref{fig:eigen-value_state} demonstrates the bulk-edge correspondence for the system without $\mathcal{PT}$ symmetry, which confirms that $\mathcal{PT}$ symmetry is not essential to the bulk-edge correspondence studied in this work.
\begin{figure}[t]
\begin{center}
\includegraphics[width=8cm]{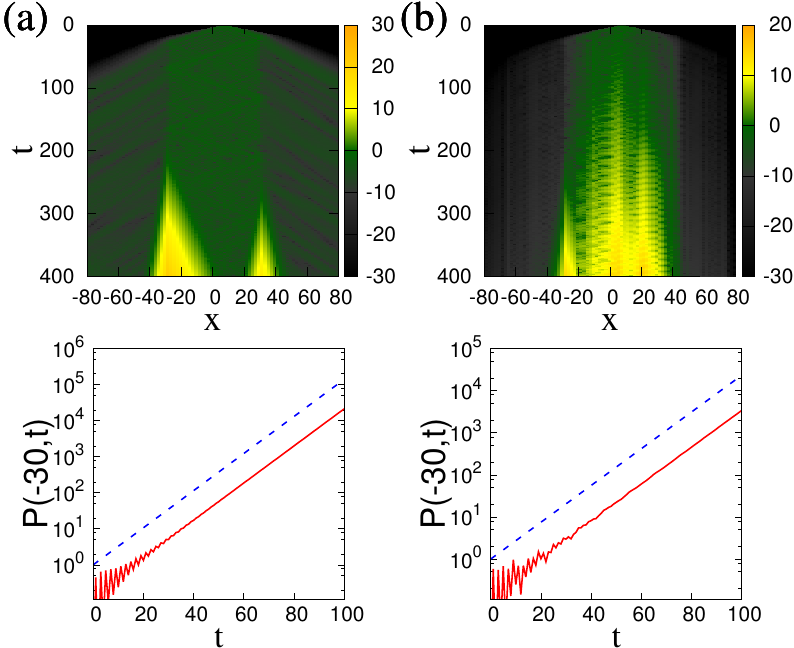}
\caption{(Color online) Dynamics of $P(x,t)$ where parameters are $\theta_1^\text{(o)}=\theta_1^\text{(i)}=9\pi/10$, $\theta_2^\text{(o)}=\theta_2^\text{(i)}=3\pi/5$, $\theta_3^\text{(i)}=\pi/5$, $\theta_3^\text{(o)}=9\pi/10$, and $L=30$, with (a) $w_j=w_\delta=0$ and (b) $w_j/\pi=w_\delta=3/20$. The parameter set is the same as that of Fig. \ref{fig:eigen-value_state} (a), excepting $w_j$ and $w_\delta$ in (a). Top row: The contour map of $\ln P(x,t)$ in the position and time step plane with $\ket{\psi(0)}=\ket{6}\otimes\ket{{\cal L}}$. Since long time evolution of 150 steps is experimentally realized in a quantum walk based on optical fiber loops \cite{regensburger2012parity}, such amplifications would be observed within the current experimental techniques. Bottom row: The semi-logarithmic plot of $P(-30,t)$ when the initial state is put near an interface $\ket{\psi(0)}=|-30\rangle\otimes\ket{{\cal L}}$. The blue dashed lines represent Eq.\ (\ref{eq:P_e}) with (a) $\kappa\approx\ln(1.06)$ and (b) $\kappa\approx\ln(1.05)$.} 
\label{fig:time-evolution}
\end{center}
\end{figure}

\subsection{Dynamics}
\label{subsec:dynamics}
Finally, we study dynamics of wavefunctions driven by the time-evolution operator $U^\prime$. From Fig. \ref{fig:eigen-value_state}, we can understand that the eigenstate with the largest $|\lambda|$ is one of topologically protected edge states. This suggests that topologically protected edge states would mainly contribute to the amplification of
\begin{align}
P(x,t)\equiv\sum_s|\psi_{x,s}(t)|^2,
\label{eq:P}
\end{align}
which corresponds to the corrected probability to find a photon in post-selected quantum systems \cite{xiao2017observation} or the intensity of light in classical systems with gain and loss  \cite{regensburger2012parity} at a site $x$ and a time step $t$.\\\indent
Figure \ref{fig:time-evolution} shows the time evolution of $P(x,t)$ with and without randomness, of which the former and latter respectively correspond to (b) and (a). When the randomness is absent and $\mathcal{PT}$ symmetry of $U^\prime$ is preserved, while almost all of eigenvalues are on the unit circle owing to $\mathcal{PT}$ symmetry, eigenvalues of edge states deviate from the unit circle. The reason why only edge states have complex quasienergy even when $w_j=w_\delta=0$ is that each edge state does not preserve $\mathcal{PT}$ symmetry due to the localization at a boundary. While eigenstates other than topologically protected edge states can break $\mathcal{PT}$ symmetry in general (see Appendix A for more details), only edge states have complex quasienergies with the parameter set in Fig. \ref{fig:time-evolution} (a). In this case, the top panel of Fig. \ref{fig:time-evolution} (a) clearly shows that $P(x,t)$ at two interfaces increase with increasing time steps, although the initial state $\ket{\psi(0)}=\ket{6}\otimes\ket{{\cal L}}$ is far from the interfaces. This is one of peculiar phenomena in $\mathcal{PT}$ symmetric nonunitary quantum walks, since observation of edge states in unitary quantum walks requires that the initial state should be very close to the interface \cite{kitagawa2012observation,barkhofen2017measuring}. On the other hand, when the randomness is induced and $\mathcal{PT}$ symmetry is broken, signals of topological edge states are difficult to see due to the amplification of other states, as shown in the top panel of Fig. \ref{fig:time-evolution} (b). Thereby the existence or absence of $\mathcal{PT}$ symmetry crucially affects the dynamics. In the bottom of Fig.\ \ref{fig:time-evolution}, we show the corrected probability near an interface $P(-30,t)$ with $\ket{\psi(0)}=|-30\rangle\otimes\ket{{\cal L}}$ and find that $P(-30,t)$ increases exponentially with time steps even when there is randomness. Taking Eqs.\ (\ref{eq:time-evolution}) and (\ref{eq:quasienergy}) into account, this enhancement of corrected probabilities is the manifestation of edge states with largest $\text{Im}(\varepsilon)$. The dashed line in Fig.\ \ref{fig:time-evolution} (b) showing
\begin{align}
P_\text{e}(t)\propto \exp(2 \kappa t),\ \kappa=\text{max}[\text{Im}(\varepsilon)]
\label{eq:P_e}
\end{align} 
confirms that the manifestation originates from $\mathcal{PT}$ symmetry breaking of the edge states.

\section{discussion and conclusion}
\label{sec:discussion_conclusion}
We have studied Floquet topological phases driven by nonunitary time evolution which satisfies chiral symmetry. We have established a procedure to calculate topological numbers in chiral symmetric open Floquet systems in Eq. (\ref{eq:nu}), based on discussions about the bulk-edge correspondence. To our knowledge, this is the first study which systematically clarifies features resulting from chiral symmetry. While the method has been applied to nonunitary Floquet systems \cite{xiao2017observation,
xiao2018higher,
chen2018characterization,
kawasaki2020bulk} based on the analogy to unitary Floquet systems \cite{asboth2013bulk}, our study gives the microscopic foundation for the validity of the procedure in nonunitary open Floquet systems. We have constructed a model classified into class BDI$^\dagger$ or AIII depending on parameters, with $\mathcal{PT}$ symmetry in some situations. Using the model, we have confirmed that the topological numbers which we have derived correctly predict the numbers of edge states. While we have computed winding numbers based on Bloch theory since the skin effect is absent in the present case, non-Bloch winding numbers can also be used if they satisfy Eqs. (\ref{eq:request_p}) and (\ref{eq:request_pp}) in the case that the skin effect occurs. Then, as a future work, it should be interesting to explore the bulk-edge correspondence based on Eq. (\ref{eq:nu}) when the skin effect is present and Bloch winding numbers are not applicable. We have also shown that topological edge states crucially affect the dynamics since they break $\mathcal{PT}$ symmetry and contribute to amplification of intensities. 

While similar systems to our model in Eq. (\ref{eq:U}) have been treated in Refs. \cite{xiao2017observation,chen2018characterization} without mentioning chiral symmetry in Eq. (\ref{eq:chiral-symmetry_each}), their topological features originate from chiral symmetry in Eq. (\ref{eq:chiral-symmetry_each}). In particular, a post selected quantum optical system in which amplifications of edge states were observed \cite{xiao2017observation} is classified into class BDI$^\dagger$ and the experimental outcomes can be understood as phenomena peculiar to $\mathcal{PT}$ and chiral symmetric open Floquet systems. The phenomena which we have shown can also be investigated from the time-step dependence of light intensity in the experimental settings of classical coherent light \cite{regensburger2012parity}. Classical systems have several advantages in comparison to quantum systems \cite{xiao2017observation} from the viewpoint of controlling open systems because time steps can be larger and gain effects can be introduced in classical systems, and so on \cite{regensburger2012parity}.

\begin{acknowledgements}
We thank Y.\ Asano and K.\ Yakubo for helpful discussions. This work was supported by KAKENHI (Grants No.\ JP18J20727, No.\ JP19H01838, No. JP18H01140, No.\ JP18K18733, and No.\ JP19K03646) and a Grant-in-Aid for Scientific Research on Innovative Areas (KAKENHI Grants No. JP15H05855 and No.\ JP18H04210) from the Japan Society for the Promotion of Science.
\end{acknowledgements}

\bibliographystyle{apsrev4-1}
\bibliography{reference.bib}

\appendix

\begin{figure}[bt]
\includegraphics[width=8cm]{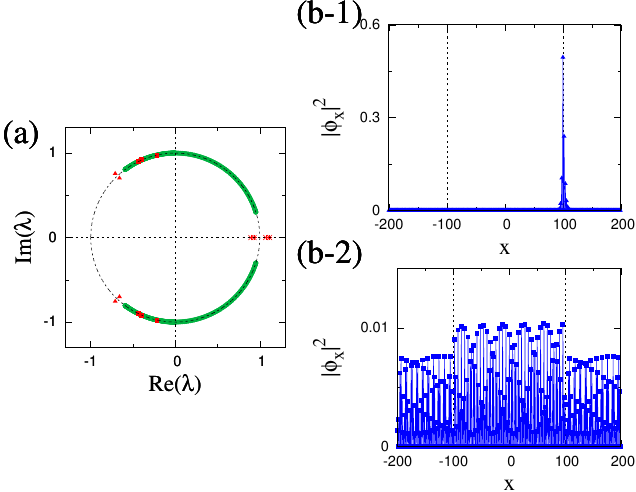}
\caption{(Color online) (a) The eigenvalues and (b) squares of wave function amplitudes of two  extra broken states for the system with the size $M=200$ and parameters $\theta_1^\text{(i)}=9\pi/10$, $\theta_2^\text{(i)}=3\pi/5$, $\theta_1^\text{(o)}=-4\pi/5$, $\theta_2^\text{(o)}=9\pi/10$, and $e^\delta=1.1$ with $\theta_3^\text{(i)}=\theta_3^\text{(o)}=w_j=w_\delta=0$. (a) Green dense dots and red symbols represent the eigenvalues whose absolute value is one and not equal to one (quasienergies are complex), respectively. More precisely, red crosses, triangles, and rectangles correspond to edge states originating from Floquet topological phases,  extra broken states inside a band gap which are localized near boundaries, and  extra broken states in the bulk spectrum which are extended throughout the system, respectively. (b) The squares of wave function amplitudes of one of extra broken states whose eigenvalue corresponds to (b-1) a red triangle and (b-2) a red rectangle in (a).}
\label{fig:eigen_value-state_detail}
\end{figure}

\section{Unbroken or broken $\mathcal{PT}$ symmetry phases for all eigenstates}
\label{sec:other broken-states}
In the case of Fig.\ \ref{fig:time-evolution} (a), only edge states with $\text{Re}(\varepsilon)=0,\pi$ break $\mathcal{PT}$ symmetry and have complex quasienergies, while the quasienergies of all the other bulk states are real due to $\mathcal{PT}$ symmetry. However, depending on the values of parameters, other eigenstates with $\text{Re}(\varepsilon) \ne 0$ or $\pi$ can have complex quasienergies even when $\mathcal{PT}$ symmetry of the time-evolution operator in Eq.\ (\ref{eq:cnd:PT:theta}) is satisfied. We refer to these states as extra broken states to distinguish them from edge states originating from nonunitary Floquet topological phases and explain details of these numerical results. \\\indent
Figure \ref{fig:eigen_value-state_detail} (a) shows eigenvalues of the nonunitary quantum walk in Eq.\ (\ref{eq:U}) satisfying Eq.\ (\ref{eq:cnd:PT:theta}). The specific parameters are as follows:
\begin{align*}
e^{\delta}=1.1,\ \theta_3^\text{(i)}&=\theta_3^\text{(o)}=w_j=w_\delta=0,\\
(\theta_1^\text{(i)}, \theta_2^\text{(i)},\theta_1^\text{(o)}, \theta_2^\text{(o)})
&=(9\pi/10,3\pi/5,-4\pi/5,9\pi/10).
\end{align*}
As shown in Fig\ \ref{fig:eigen_value-state_detail} (a), eigenvalues corresponding to the extra broken states shown by red symbols appear not only within a band gap but also in the bulk spectra.
The former eigenstates shown in Fig.\ \ref{fig:eigen_value-state_detail} (b-1) are localized near boundaries, while the latter eigenstates shown in Fig.\ \ref{fig:eigen_value-state_detail} (b-2) are extended in the whole system. However, we emphasize that imaginary parts of complex quasienergies of edge states originating from nonunitary Floquet topological phases are larger than those for the extra broken states, which we can understand from Fig.\ \ref{fig:eigen_value-state_detail} (a). Thereby, when we consider the time evolution for the system even with extra broken states, edge states originating from Floquet topological phases dominate the exponential amplification of intensity at the interfaces.

\end{document}